\documentstyle[prl,aps,amsfonts,amssymb,twocolumn,epsfig]{revtex}
\begin{document}
\draft

\newcommand{\w}{\omega}
\newcommand{\A}{\text{\AA}}
\newcommand{\IM}{\text{Im}}
\renewcommand{\vec}[1]{{\bf #1}}
\wideabs{
\title{Two-dimensional fluctuations at the quantum-critical point of 
CeCu$_{6-x}$Au$_x$}
\author{O. Stockert$^1$, H. v. L\"ohneysen$^1$, A. Rosch$^2$, N. Pyka$^3$
and M. Loewenhaupt$^4$}
\address{$^1$Physikalisches Institut, Universit\"at Karlsruhe, 
D-76128 Karlsruhe,
Germany\\
$^2$Institut f\"ur Theorie der Kondensierten Materie, Universit\"at Karlsruhe, 
D-76128 Karlsruhe, Germany\\
$^3$Institut Laue-Langevin, F-38042 Grenoble, France\\
$^4$Institut f\"ur Angewandte Physik, Technische Universit\"at Dresden, 
D-01062 Dresden, Germany}
\date{\today}
\maketitle

\begin{abstract}
The heavy-fermion system CeCu$_{6-x}$Au$_x$ exhibits a quantum critical
point at $x_c \approx 0.1$ separating nonmagnetic and magnetically ordered
ground states. The pronounced non-Fermi-liquid behavior at $x_c$ calls for a
search for the relevant quantum critical fluctuations. Systematic
measurements of the inelastic neutron scattering cross section $S (q,\w)$
for $x = 0.1$ reveal rod-like features in the reciprocal $ac$ plane
translating to two-dimensional ($2d$) fluctuations in real space. We find
$3d$ magnetic ordering peaks for $x = 0.2$ and $0.3$ located on these rods
which hence can be viewed as $2d$ precursors of the $3d$ order.
%We have performed neutron scattering on the heavy-fermion system
%CeCu$_{6-x}$Au$_x$ which exhibits a quantum critical point at $x_c \approx
%0.1$ separating nonmagnetic and magnetically ordered ground states. The
%non-Fermi-liquid behavior observed at $x_c$ calls for a search for the
%relevant critical fluctuations. A detailed investigation of the inelastic
%scattering cross section $S (q,\w)$ for $x = 0.1$ reveals rod-like features
%in the reciprocal $ac$ plane translating to two-dimensional ($2d$)
%fluctuations in real space. The $3d$ magnetic ordering peaks for $x = 0.2$
%and $0.3$ are located on these rods which hence can be viewed as $2d$
%precursors of the $3d$ order.
\end{abstract}
}

Continuous quantum phase transitions which occur in a strict sense only at
temperature $T = 0$ are driven by quantum fluctuations instead of thermal
fluctuations as for ordinary classical phase transitions
\cite{hertz76,millis93}.  This leads to unusual and rich behavior even at
finite temperatures in the neighborhood of the critical point. Because of the
uncertainty principle the energy scale of fluctuations introduces a time scale
which leads to an intricate coupling of static and dynamic critical behavior.
For instance, the critical behavior of the specific heat will depend on the
dynamical critical exponent $z$ relating the typical lifetime $\xi_{\tau}$ and
correlation length $\xi$ of critical fluctuations, $\xi_{\tau} \propto
\xi^{z}$.  Such a quantum phase transition can be achieved by changing a
coupling parameter which plays a role analogous to temperature in ordinary
phase transitions.  In recent years, many physical realizations of quantum
phase transitions have been found. The case of a magnetic-nonmagnetic
transition in heavy-fermion metals is particularly interesting because of the
involvement of itinerant electrons.

Excitations of a system of interacting itinerant 
electrons in a metal, i.e. quasiparticles, are usually described within the
Fermi-liquid theory, with the specific heat $C \propto T$, a Pauli 
susceptibility
independent of $T$, and an electrical resistivity contribution $\Delta \rho
\propto T^2$ due to quasiparticle-quasiparticle scattering. Interactions 
renormalize the quasiparticle masses with respect to the free-electron mass
$m_0$. Even in heavy-fermion systems with quasiparticle masses as high as
several 100 $m_0$,
Fermi-liquid behavior is the rule rather than the exception
\cite{grewe}. In heavy-fermion systems, the coupling parameter tuning the 
magnetic-nonmagnetic transition is the (antiferromagnetic) exchange interaction
$J$ between $4f$ or $5f$ magnetic moments and conduction electrons 
\cite{grewe}.
If it is strong, a local singlet state is formed via the Kondo effect around
each $4f$ or $5f$ site, leading to a nonmagnetic ground state. On the other
hand, a weak (but non-zero) exchange interaction leads to a 
Rudermann-Kittel-Kasuya-Yosida coupling between moments and hence to magnetic
order. In the exemplary system CeCu$_{6-x}$Au$_x$ doping of CeCu$_6$ with 
the larger Au atom leads - via lattice expansion - 
to a weakening of the Kondo effect and hence to 
long-range antiferromagnetic order for $x > x_c \approx 0.1$, with a 
linear increase of the N\'eel temperature $T_N \propto (x - x_c)^{\mu}$,
i.e. $\mu = 1$ \cite{loehneysen96}.

At $x_c$ where $T_N$ vanishes, i.e. around the quantum critical point, 
pronounced deviations from Fermi-liquid behavior occur. This non-Fermi-liquid
(NFL) behavior is seen, e.g. in the specific heat where $C/T \propto {\rm -ln}
(T/T_0)$ over nearly two decades and in the resistivity where
$\Delta \rho \propto T$. It is precisely this NFL 
behavior at the quantum
critical point that stirred a lot of interest \cite{nfl} since it cannot
be explained in terms of a transition driven by three-dimensional $(3d)$ 
fluctuations,
because for an antiferromagnet with $d = 3$ and $z = 2$, 
$C/T \propto  1 - B \surd T$
and $\Delta \rho \propto T^{3/2}$ would be expected \cite{millis93,moriya95}.

A step forward towards the solution of the NFL puzzle in CeCu$_{6-x}$Au$_x$
was to realize \cite{rosch97} that $2d$ critical fluctuations coupled 
to quasiparticles with $3d$ dynamics will indeed lead to $\gamma = C/T 
\propto {\rm -ln}(T/T_0)$, $\Delta \rho \propto T$ and $\mu = 1$ as 
experimentally observed. Elastic
neutron scattering experiments at 0.07\,K on CeCu$_{5.8}$Au$_{0.2}$ with
$T_N$ = 0.25\,K showed, in addition to peaks attributed to (short-range)
antiferromagnetic order, broad maxima along the $a^*$ direction that were
much sharper in the $b^*$ direction \cite{rosch97}. This latter feature was interpreted 
in terms of ferromagnetic planes perpendicular to the $a$ direction 
(orthorhombic notation) and thus provided a possible scenario of the 
$d = 2$, $z = 2$ universality class \cite{rosch97}.
Without any direct evidence it is certainly hard to believe that $2d$
correlations are dominating an intrinsically $3d$ alloy, even if the 
thermodynamics strongly support such a picture. Therefore it is essential 
to investigate the quantum critical fluctuations directly by inelastic
neutron scattering.

\begin{figure}[h]
\centering
\epsfxsize=6.5cm
\mbox{\epsfclipon\epsfbox{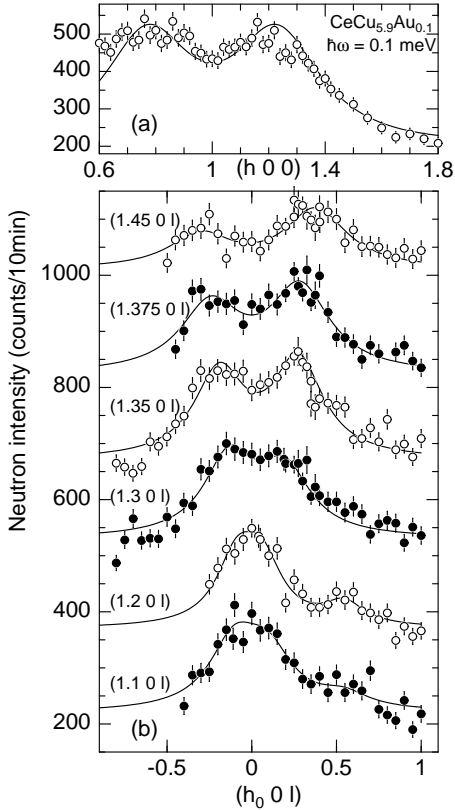}}
\caption[]{a) $q$-scan along [1~0~0] in CeCu$_{5.9}$Au$_{0.1}$ 
            at $T = 50$\,mK.
           b) $q$-scans $(h_0~0~l)$ in the $a^*c^*$-plane in the $c^*$
            direction for fixed $h_0$ at $T = 70$\,mK (all scans with fixed
            $k_f = 1.15$\,{\AA}$^{-1}$ and $\hbar\w = 0.1$\,meV). The scans
            are shifted by 150 counts to each other.}
\label{fig1}
\end{figure}

The experiments were carried out at the triple-axis spectrometer IN14 at the
Institut Laue-Langevin, Grenoble with a fixed final neutron energy 
$E_f = 2.7$\,meV ($k_f = 1.15 \A^{-1}$), giving an energy resolution (FWHM) 
of 0.07\,meV. The 
CeCu$_{6-x}$Au$_x$ single crystals were grown
with the Czochralski method in a W crucible. The specific heat of the sample
with $x = 0.1$ exhibits the NFL behavior $C/T \propto {\rm -ln}(T/T_0)$ as 
measured down to 60\,mK, in agreement with previous samples of the
same Au concentration \cite{loehneysen96}.

Fig.\,1 shows $q$ scans of the dynamic structure factor $S(q,\omega)$
in the reciprocal $ac$ plane in two perpendicular 
directions at very low $T < 100$\,mK at an energy transfer 
$\hbar \omega = 0.1$\,meV. The
(h 0 0) scan (Fig.\,1a) reveals a broad double maximum at (0.8 0 0) and 
(1.2 0 0). This double maximum is only resolved at small $\hbar \omega$.
For instance, for $\hbar \omega = 0.25$\,meV only a single broad feature
centered at (1 0 0) is seen \cite{stockert97}. Hence it may be thought of as 
developing from the broad maximum observed at the same $q$ for $\hbar \omega$
= 0.3\,meV in CeCu$_6$ \cite{rossat88}. Upon entering the magnetically
ordered state for the $x = 0.2$ alloy, the double-peak structure appears
as a (quasi-)elastic feature for $x = 0.2$ that 
represents short-range ordering evidenced by a width in $q$ that is considerably 
larger than the $q$ resolution \cite{rosch97,stockert97}. 

Fig.\,1b shows that for $x = 0.1$ there is a very rich structure of 
$S(q,\omega)$ in the $a^*c^*$ plane,
as derived from scans $(h_0~0~l)$ along $c^*$ for fixed $h = h_0$. The peak at 
(1.2 0 0) splits when moving away from the $a^*$ axis. The solid lines
present Lorentzian fits with a width of $(0.24 \pm 0.02)$ {\AA}$^{-1}$
for all scans shown. However, the main point is that
the peak height remains roughly constant across the whole Brillouin zone 
(cf Fig.\,3a). The width along $c^*$ is comparable to the width of the 
(1.2 0 0) maximum along $(h~0~0)$ (cf.
Fig.\,1a). This suggests a rod-like feature of the 
dynamical magnetic response. It does not, however, extend along the
$a^*$ axis as previously assumed \cite{rosch97} but in an oblique direction.

\begin{figure}[h]
\centering
\epsfxsize=6.5cm
\mbox{\epsfclipon\epsfbox{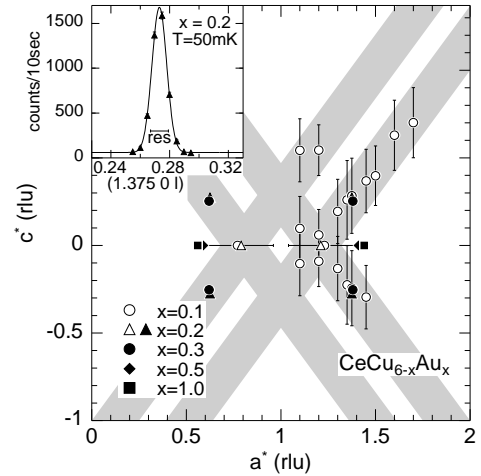}}
\caption[]{Position of the dynamic correlations ($x = 0.1$, 
           $\hbar\w = 0.1$\,meV, $T < 100$\,mK)
           and magnetic Bragg peaks ($x = 0.2 - 1.0$) in the reciprocal
           $ac$ plane in CeCu$_{6-x}$Au$_{x}$ (data for $x =$ 0.3 and 1.0 from
           \cite{stockert}, for $x = 0.5$ from \cite{schroeder94}). The 
           vertical
           and horizontal bars indicate the line width for $x = 0.1$. The four
           rods are related by the orthorhombic symmetry (Pnma). Here we
           ignore a small monoclinic distortion ($< 1 \%$) at low temperatures.
           The open symbols for $x = 0.2$ represent
           the short-range ordering peaks \cite{rosch97,stockert97}. The inset shows a resolution
           limited magnetic Bragg peak for $x = 0.2$ at $T = 50$\,mK 
           (elastic scan at $(1.375~0~l$)).}
\label{fig2}
\end{figure}

Fig.\,2 shows the peak positions in the $a^*c^*$ plane for $x = 0.1$ and 
$\hbar \omega$ = 0.1\,meV derived from Fig.\,1 and from further
measurements. To corroborate the
rod-like nature of $S(q,\hbar\omega = 0.1\,{\rm meV}$) for $x = 0.1$,
further scans across the peaks were performed along independent directions, 
one in the $a^*b^*$ plane along $b^*$ and one in the $a^*c^*$ plane 
perpendicular to the rod-like feature.
They, too, reveal a width of comparable magnitude as can be seen from Fig.\,3b
and c.
In order to interpret $S(q,\omega)$ of CeCu$_{5.9}$Au$_{0.1}$, we recall 
that a rod-like feature in $q$ space is related to a $2d$
correlation between Ce atoms in real space. The shaded 
rods in Fig.\,2 can therefore be identified with planes in real space. These 
planes extend into the $b$ direction and into a direction in the $ac$ plane
given by next-nearest neighboring Ce atoms.
Thus, the observed quasi
$2d$ correlations strongly support the proposed scenario \cite{rosch97}
of $2d$ spin fluctuations coupled to quasiparticles with 
$3d$ dynamics, although the quasi $2d$ correlations are 
not ferromagnetic as initially supposed.

The $2d$ fluctuations apparently are the precursor of the $3d$ magnetic
ordering. Indeed, the Bragg points for samples not too far from the magnetic
instability, e.g. $x = 0.2$ and 0.3, are located on the rods for $x = 0.1$.
For $x = 0.2$ in addition to the rather broad double maximum at $q$ = (0.8 0
0) and (1.2 0 0) \cite{rosch97} we find resolution-limited peaks at (0.625 0
0.275) and at lattice-equivalent positions in reciprocal space. The inset in
Fig.\,2 displays a $(1.375\,0\,l)$ scan of such a Bragg peak. The main frame
of Fig.\,2 shows that its position is indeed on one of the rods as is the
position of the short-range order peaks along $a^*$. However, we have {\em not}
observed a $3d$ precursor for $x = 0.1$, i.e. enhanced scattering intensity
around the Bragg peak for $x = 0.2$. This is an important point in favor of
the $2d$ scenario. For $x = 0.3$, the Bragg position remains almost unchanged
while no short-range order peaks on the $a^*$ axis were detected. For $x =
0.5$ a sudden reorientation of the magnetic ordering vector is observed, with
incommensurate order along $a^*$ with $\bf \tau$ = (0.59 0 0)
\cite{schroeder94} which is then roughly constant up to $x = 1$
\cite{stockert}. The reorientation of $\bf \tau$ occurring between $x = 0.3$
and 0.5 deserves further study.

Returning to the quantum-critical point at $x=0.1$, we recall that in
the $d=2$, $z=2$ scenario we expect the following generic form of the
$\w$-- and $q$-dependent susceptibility describing magnetic
fluctuations in the plane \cite{rosch97}:
\begin{eqnarray}
  \label{chi}
  \chi^{2d}_{\vec{q}_{\|}}(\w) \approx \mu_B^2 \left[\frac{1}{q_0^2} 
\left(\frac{1}{\xi^2}+\vec{q}_{\|}^2\right)- i \w/\w_0\right]^{-1}
\end{eqnarray}
The imaginary part of the susceptibility is directly proportional to
the magnetic structure factor $S(\vec{q},\w)$ measured with inelastic
neutron scattering, $S(\vec{q},\w)=(1+n_B(\w)) \IM
\chi^{2d}_{\vec{q}_{\|}}(\w) f(q_{\perp})$, where $n_B(\w)$ is the
Bose-function. The smooth function $f(q_{\perp})$ describes the weak
$q$-dependence perpendicular to the planes, i.e. along the
rod-like structures shown in Fig.~\ref{fig2}.  $\vec{q}_{\|}$ is the
momentum in the plane, i.e.  perpendicular the rods.  $q_0$ and
$\omega_0$ are constants that vary only slightly with temperature $T$ and
depend only weakly on the momentum along the rods. We expect that the
above equation is valid (up to logarithmic corrections) for small
momenta and frequencies $\w, \w_0 \vec{q}_{\|}^2/q_0^2 \ll k_B T_K$,
where $T_K \approx 6$\,K is the Kondo temperature in the system. We have
neglected the small anisotropy within the planes. The effective
correlation length is given by $\xi$. It is expected to vary strongly
with $T$, $\w_0/(q_0 \xi)^2 \approx A k_B T$ with $A$ being
a constant of order one varying only logarithmically with $T$
\cite{sachdev,millis93}.

The $q$-scans were performed with an energy transfer of $0.1 $\,meV 
($=  1.2 {\rm K} \cdot k_{\rm{B}}$) which is large
compared to the temperature of $70$\,mK. Therefore we expect that the width
\begin{figure}[h]
\centering
\epsfxsize=6.5cm
\mbox{\epsfclipon\epsfbox{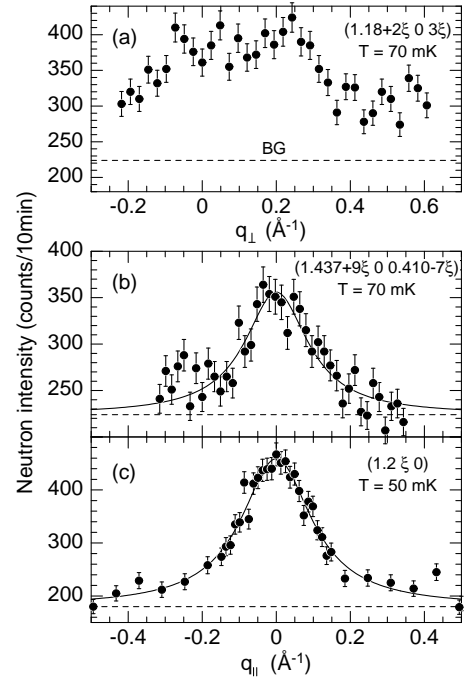}}
\caption[]{$q$-scans along (a) and across (b, c) the rod-like feature in 
           CeCu$_{5.9}$Au$_{0.1}$ for $\hbar\w = 0.1$\,meV and 
           $k_f = 1.15$\,{\AA}$^{-1}$ ($T = 50/70$\,mK). There is only
           a weak $q$-dependence along the rods ($q_\perp$), while transverse
           scans ($q_{\|}$) show peaks with nearly the same line width.} 
\label{fig3}
\end{figure}
of the peaks shown in Fig.~\ref{fig1} and \ref{fig3} does {\em not} 
measure the true correlation length $\xi$ but determines the ratio 
\begin{equation}
q_0^2/\w_0 \approx (0.1 \pm 0.02) (\text{meV \AA})^{-1}
\end{equation}

The important question arises whether the observed magnetic
fluctuations can be related to the NFL behavior of the
thermodynamic quantities at the quantum critical point, i.e. to the
logarithmically diverging specific-heat coefficient. Actually, the
prefactor of the specific-heat coefficient per area of a {\em plane}
is fully determined by the quantum critical theory \cite{vorfaktor}
$\gamma^{2d}=(n/12) (q_0^2/\w_0) \ln (T_0/T)$ where $n$ is the
number of spin components. We use $n=1$, as the magnetic anisotropy
\cite{loehneysen96}  suggests an Ising system. $T_0$ is an unknown
temperature scale of the
order of $T_K$. To calculate the specific heat per volume one has to
know the distance $L$ of the planes -- note that especially in an
incommensurate structure such a distance is only an effective
quantity.  Then the molar specific heat is given by
\begin{equation}\label{gammaM}
\gamma \approx  2 \frac{V_{M}  \gamma_{\text{plane}}}{L} \ln (T_0/T)
\approx
 (1.3\pm 0.2) \frac{\text{J}}{\text{mol K}^2} \frac{\A}{L} \ln (T_0/T)
\end{equation}
$V_M$ is the volume per mol of CeCu$_{5.9}$Au$_{0.1}$, the factor $2$
takes into account that the correlations show up in two different
directions (see Fig.~2). This value has to be compared to the measured 
\cite{loehneysen96} specific heat coefficient $\gamma= 0.6\,
\frac{\text{J}}{\text{mol K}^2} \ln (T_0/T)$ which is indeed of the same
order of magnitude as our estimate (\ref{gammaM}).  From the crystal
structure one would expect $L$ to be of the order of $4$ to $10\A$
while from (\ref{gammaM}) we obtain $L\approx 2-3\A$ which is somewhat
too small.  However, one has to take into account the considerable
theoretical uncertainty, e.g. arising from the definition of $L$ in an
incommensurate system or the unknown effective number of spin-components
in this anisotropic system. In addition, the momenta and frequencies used 
in our analysis are quite large.

 Therefore, we think that the
semi-quantitative agreement of the width of the rods in $q$ space
compared to the specific heat gives strong support for the idea that the
$2d$ fluctuations are responsible for the observed NFL behavior.

\begin{figure}[h]
\centering
\epsfxsize=6.5cm
\mbox{\epsfclipon\epsfbox{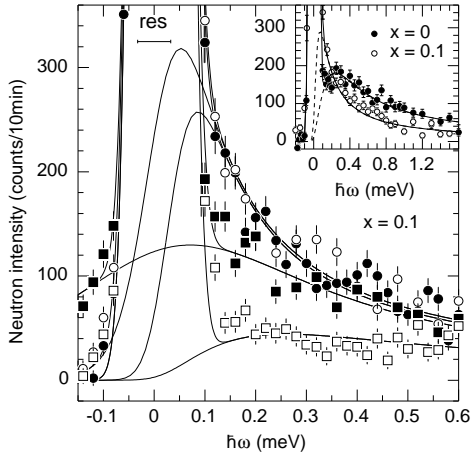}}
\caption[]{Energy scans at $q$ = (1.375~0~0.275) in 
           CeCu$_{5.9}$Au$_{0.1}$ for $k_f = 1.15$\,{\AA}$^{-1}$ for
           different temperatures: $\circ$ 70\,mK, $\bullet$ 
           500\,mK and {\tiny $\blacksquare$} 5\,K. For comparison 
           an energy scan 
           at $q$ = (1.8~0~0) for $T = 70$\,mK is displayed 
           ({\tiny $\square$}). 
           The inset shows energy scans at
           (1.2~0~0) in  CeCu$_{6-x}$Au$_{x}$ for $x = 0$ ($T = 80$\,mK)
           and  $x = 0.1$ ($T = 50$\,mK),  $k_f = 1.15$\,{\AA}$^{-1}$. 
           The background is subtracted for all scans.}
\label{fig4}
\end{figure}

As a final point, we discuss the energy dependence of the critical modes.
Fig.\,4 shows energy scans at exactly the $q$ value of the magnetic order in 
CeCu$_{5.8}$Au$_{0.2}$ (cf. inset of Fig.\,2). The solid lines indicate 
a fit comprised of an elastic Gaussian and a quasi-elastic Lorentzian (convoluted with
the resolution) with
full-width $\Gamma = 0.62, 0.14$ and 0.11\,meV for the latter at 
$T = 5, 0.5$ and 0.07\,K, respectively. The shift towards $\omega = 0$
with decreasing $T$ is clearly seen at this $q$ as opposed to the rather
broad $\omega$ scan at $q = (1.8\,0\,0)$ with $\Gamma$ = 0.48\,meV for $T$ =
0.07\,K (deduced from polycrystalline measurements). Due to the finite energy
resolution and the large elastic background it is difficult to obtain
reliable values for $\Gamma$ at low temperatures. For instance, assuming
a weakly $\w$-dependent background reduces $\Gamma$ at $T = 0.07$\,K by a 
factor of 2-3. Nevertheless, the observed decrease of $\Gamma$ seems to
be slower than $\Gamma = \w_0/(q_0 \xi)^2 \propto T$ predicted for  the 
$d = 2$, $z = 2$ quantum critical theory for $T \ll T_K \approx$ 6\,K. It is
important to note that the onset of $3d$ correlations would have the opposite 
effect at this $q$ value where $3d$
magnetic order is expected. We note that in measurements of $S(q,\w)$ 
concentrated on $q$ = (1.2 0 0) a similar decrease of $\Gamma$ towards
low $T$ with apparent leveling-off is observed \cite{schroeder}.
On the other hand, the thermodynamics, notably the specific heat, shows a 
logarithmic increase of $\gamma$ down to at least 0.06\,K without a
signature of a corresponding energy scale. Therefore this point certainly
requires further studies, e.g. on the role of disorder in our
incommensurate system.

In conclusion, we have identified the critical two-dimensional fluctuations 
leading to non-Fermi-liquid behavior in CeCu$_{5.9}$Au$_{0.1}$ with a 
systematic study of quasielastic neutron scattering. From the observed dynamic
susceptibility a semi-quantitative agreement with the prefactor of the 
logarithmic increase of the specific-heat coefficient is found.

We acknowledge helpful discussions with A. Schr\"oder and P. W\"olfle. This
work was supported by the Deutsche Forschungsgemeinschaft.


\begin{references}
\bibitem{nfl} NFL behavior in heavy-fermion systems can also occur because
of single-ion effects, e.g. the quadrupolar Kondo effect or a distribution of
Kondo effects. These mechanisms are not operative in CeCu$_{6-x}$Au$_x$.
\bibitem{vorfaktor}
The formula given here is a factor of $1/8$ smaller than the result given
by Millis \cite{millis93} correcting small errors.
\end{references}
\end{document}